\documentstyle[11pt,aaspp4]{article}

\def\la{\mathrel{\hbox{\rlap{\hbox{\lower4pt\hbox{$\sim$}}}\hbox{$<$}}}}
\def\ga{\mathrel{\hbox{\rlap{\hbox{\lower4pt\hbox{$\sim$}}}\hbox{$>$}}}}

\slugcomment{Accepted to {\it The Astrophysical Journal}}

\begin{document}

\title{Constraints on the Redshift and Luminosity Distributions of Gamma Ray 
Bursts in an Einstein-de Sitter Universe}

\author{Daniel E. Reichart\altaffilmark{1,2} and P. M\'esz\'aros\altaffilmark{1,3}}

\altaffiltext{1}{Department of Astronomy and Astrophysics, Pennsylvania State University, University Park, PA 16802}
\altaffiltext{2}{Department of Astronomy and Astrophysics, University of Chicago, Chicago, IL 60637} 
\altaffiltext{3}{Center for Gravitational Physics and Geometry, Pennsylvania State University, University Park, PA 16802} 

\begin{abstract}
Two models of the gamma ray burst population, one with a standard candle 
luminosity and one with a power law luminosity distribution, are 
$\chi^2$-fitted to the union of two data sets:  the differential number 
versus peak flux distribution of BATSE's long duration 
bursts, and the time dilation and energy shifting versus peak flux information 
of pulse duration time dilation factors, interpulse duration time dilation 
factors, and peak energy shifting factors. The differential peak flux 
distribution is corrected for threshold effects at low peak fluxes and 
at short burst durations, and the pulse duration time dilation factors are also 
corrected for energy stretching and similar effects. Within an Einstein-de 
Sitter cosmology, we place strong bounds on the evolution of the bursts, 
and these bounds are incompatible with a homogeneous population, assuming a 
power law spectrum and no
luminosity evolution. Additionally, under the implied conditions of moderate 
evolution, the 90\% width of the {\it observed} luminosity distribution is 
shown to be $\la$ 10$^2$, which is less constrained than others have 
demonstrated it to be assuming no evolution.
Finally, redshift considerations indicate that if the redshifts of BATSE's 
faintest bursts are to be compatible with that which is currently known for 
galaxies, a standard candle luminosity is unacceptable, and in the case
of the power law luminosity distribution, a mean luminosity $\la$ 10$^{57}$
ph s$^{-1}$ is favored.

\end{abstract}

\keywords{gamma-rays: bursts - cosmology: theory}

\section{Introduction}

The angular distribution of the gamma ray burst population has been shown to 
be highly isotropic (\cite{mea92}; \cite{bea96}).  This suggests that the 
bursts are either located in an extended galactic halo (e.g., \cite{p91}) or 
that they are cosmological in origin (e.g., \cite{p86}).  Recent measurements 
of time dilation of burst durations (\cite{nea94}, 1995; \cite{wp94}; however, 
see \cite{mea96}), of pulse durations (\cite{nea96a}), and of interpulse 
durations (\cite{d95}; \cite{nea96b}) in the BATSE data, as well as 
measurements of peak energy shifting (\cite{mea95}), favor the latter 
explanation.  

Models, both galactic and cosmological, are typically fitted to the 
differential peak flux distribution of BATSE's long duration ($T_{90} >$ 2 s) 
bursts. Furthermore, this distribution is typically truncated at a peak flux of
1 ph cm$^{-2}$ s$^{-1}$ to avoid threshold effects.  Here, we fit two models, 
one with a standard candle luminosity and one with a power law luminosity 
distribution, to not only BATSE's 3B differential distribution, but also 
to the pulse duration time dilation factors (corrected for energy stretching 
and similar effects) of Norris et al. (1996a), the interpulse duration time 
dilation factors of Norris et al. (1996b), and the peak energy shifting factors 
of Mallozzi et al. (1995).  These three independent sets of measurements are 
shown to be self-consistent in \S4.  (All three are for long duration bursts 
only.)  
Furthermore, via the analysis of Petrosian \& Lee (1996a), BATSE's differential 
distribution is extended down to a peak flux of 0.316 ph cm$^{-2}$ s$^{-1}$, 
which corresponds to a trigger efficiency of $\sim \frac{1}{2}$ on BATSE's 1024 ms timescale.  

Together, the differential distribution and the time dilation and energy 
shifting factors place strong bounds on the evolution of the burst population.
These bounds favor
moderate evolution and are incompatible with homogeneity, assuming only minimal 
luminosity evolution.  This result is compatible with the analyses of Fenimore 
\& Bloom (1995), Nemiroff et al. (1996), and Horack, Mallozzi, \& Koshut (1996).
Furthermore, under these conditions of moderate evolution, the 90\% width of 
the {\it observed} luminosity distribution is shown to be less constrained than 
others have demonstrated it to be assuming no evolution (see \S5).  
Finally, redshift considerations indicate that 
if the redshifts of BATSE's faintest bursts are to be compatible with that
which is currently associated with the formation of the earliest galaxies,
the mean luminosity of the bursts should be $\sim$ 10$^{57}$ ph s$^{-1}$ or 
lower.

\section{Cosmological Models}

Both the standard candle luminosity model and the power law luminosity 
distribution model assume a power law redshift distribution, given by
\begin{equation}
n(z) = n_0(1+z)^{D},
\end{equation}
where $n(z)$ is the number density of bursts of redshift $z$.  This 
distribution is bounded by 0 $< z < z_M$, where $z_{M}$ is the maximum burst 
redshift. The luminosity distributions of the two models are given by 
\begin{equation}
\phi(L) = \cases{\phi_0\delta(L - L_0) & (standard candle) \cr \phi_0L^{-\beta} & (power law)}.
\end{equation}
The standard candle is of luminosity $L_0$ and the power law luminosity is 
bounded by minimum and maximum luminosities $L_m < L < L_M$.  All luminosities 
are peak photon number luminosities and all fluxes are peak photon number 
fluxes (measured over BATSE's 50 - 300 keV triggering range); however, see recent 
papers by Bloom, Fenimore, \& in 't Zand (1996) and Petrosian \& Lee (1996b) 
which introduce the fluence measure.  

\subsection{Integral Distribution}

Assuming a power law spectrum and an Einstein-de Sitter cosmology, the bursts' 
integral distribution, i.e. the number of bursts with peak fluxes greater than 
an arbitrary value $F$, is given for either model by (\cite{mm95}) 
\begin{equation}
N(>F) = \frac{32\pi n_0c^3}{H_0^3}\int_{L_m}^{L_M}\phi(L)dL\int_0^{\chi_0}(1-\chi)^{8-2D}\chi^2d\chi,
\label{3a}
\end{equation}
where
\begin{equation}
\chi_0 = min(\chi_1,\chi_2),
\end{equation}
\begin{equation}
\chi_1 =\frac{1}{1+{\frac{4c}{H_0}}\left({\frac{\pi F}{L}}\right)^{\frac{1}{2}}},
\label{3c}
\end{equation}
and
\begin{equation}
\chi_2 = 1 - \frac{1}{(1+z_M)^{\frac{1}{2}}}.
\label{3d}
\end{equation}
A photon number spectral index of -1 (or a power-per-decade spectral
index of 1) has been assumed.  This value is typical of burst spectra, 
especially at those frequencies at which most of the photons are received 
(e.g., Band et al. (1993)).
In the case of the standard candle model, eq. \ref{3a} becomes
\begin{equation}
N(>F) \propto \int_0^{\chi_0}(1-\chi)^{8-2D}\chi^2d\chi,
\label{9a}
\end{equation}
where $L = L_0$ in eq. \ref{3c}.  The factor of proportionality has been 
dropped because only normalized integral distributions (see \S 3.1) and ratios 
of integral distributions (see \S 2.2) are fit to.  Eq. \ref{9a} has the 
analytic solution
\begin{equation}
N(>F) \propto f(\chi_0,8-2D),
\label{10a}
\end{equation}
where
\begin{equation}
f(\chi,q)= \frac{2(1 - (1-\chi)^{3+q})}{(1+q)(2+q)(3+q)} -\frac{2\chi(1 - \chi)^{2+q}}{(1+q)(2+q)}-\frac{\chi^2(1-\chi)^{1+q}}{1+q}.
\end{equation}
In the case of the power law model, eq. \ref{3a} becomes
\begin{equation}
N(>F) \propto \int_1^K x^{-\beta}dx\int_0^{\chi_0}(1-\chi)^{8-2D}\chi^2d\chi,
\label{14a}
\end{equation}
where
\begin{equation}
K = \frac{L_M}{L_m}
\end{equation}
and $L = xL_m$ in eq. \ref{3c}.  Eq. \ref{14a} has the integral solution
\begin{equation}
N(>F) \propto \int_1^Kf(\chi_0,8-2D)x^{-\beta}dx.  
\label{18}
\end{equation}

\subsection{Time Dilation and Energy Shifting Factors}

In an idealized scenario of two identical bursts at different redshifts, $z_1$ 
and $z_2$, their time dilation and energy shifting factors, $\tau_{12}$ and 
$\epsilon_{12}$, are both simply equal to the ratios of their scale factors 
(neglecting the effects of energy stretching which are inherent in pulse 
duration measurements (\cite{fb95})):
\begin{equation}
\tau_{12} = \epsilon_{12}^{-1} = \frac{1 + z_1}{1+z_2}.
\end{equation}
In practice, however, measures of the scale factor are averaged over peak flux 
ranges and time dilation and energy shifting factors are determined for pairs 
of these ranges.  M\'esz\'aros \& M\'esz\'aros (1996) demonstrated that such 
mean values of the scale factor, averaged over a peak flux 
range $F_l < F < F_u$, are 
simple functions of the integral distribution, as modeled by eqs. \ref{10a} 
and \ref{18}:
\begin{equation}
\overline{(1+z)}(F_l,F_u) = \frac{N_{D+1}(F_l,F_u)}{N_D(F_l,F_u)},
\end{equation}
where 
\begin{equation}
N(F_l,F_u) = N(>F_u) - N(>F_l).
\label{diff}
\end{equation}
Consequently, time dilation and energy shifting factors between two such 
ranges, $F_{1,l} < F_1 < F_{1,u}$ and $F_{2,l} < F_2 < F_{2,u}$, are given by
\begin{equation}
\tau_{12} = \epsilon_{12}^{-1} = \frac{N_{D+1}(F_{1,l},F_{1,u})N_D(F_{2,l},F_{2,u})}{N_D(F_{1,l},F_{1,u})N_{D+1}(F_{2,l},F_{2,u})}.
\label{8}
\end{equation}
The effects of energy stretching are not modeled here because they are 
removed empirically from the pulse duration measurements of Norris et al.
(1996a) in \S3.2.  The interpulse duration measurements of Norris et al. 
(1996b) and the peak energy measurements of Mallozzi et al. (1995) do not 
require such corrections.

\section{Data Analysis}

\subsection{Integral Distribution}

BATSE's sensitivity becomes less than unity at peak fluxes below $\sim$ 1 ph 
cm$^{-2}$ s$^{-1}$ (\cite{fea93}).  
Petrosian, Lee, \& Azzam (1994) demonstrated that BATSE is additionally biased 
against short duration bursts:  BATSE triggers when the 
mean photon count rate, defined by
\begin{equation}
\bar C(t) = \frac{1}{\Delta t}\int_t^{t + \Delta t}C(t)dt
\end{equation}
where $\Delta t =$ 64, 256, and 1024 ms are BATSE's predefined timescales, 
exceeds the threshold count rate, $\bar C_{lim}$, on a particular timescale.  
Consequently, peak photon count rates are underestimated for bursts of duration 
$T \la \Delta t$, sometimes to the point of non-detection.  Peak fluxes are 
similarly underestimated.  Petrosian \& Lee (1996a) developed (1) a correction 
for BATSE's measured peak fluxes and (2) a non-parametric method of correcting 
BATSE's integral distribution.  

A burst's corrected peak flux is given by
\begin{equation}
F = \bar F\left(1 + \frac{\Delta t}{T_{90}}\right),
\label{20}
\end{equation}
where $\bar F$ is the burst's measured peak flux and $T_{90}$ is the burst's 
90$\%$ duration.  Consequently, if $T_{90} \gg \Delta t$, $F \simeq \bar F$; 
however if $T_{90} \la \Delta t$, $F > \bar F$. Petrosian \& Lee (1996a)
demonstrated 
that eq. \ref{20} adequately corrects BATSE's measured peak fluxes (1) on the 
1024 ms timescale, (2) for bursts of duration $T_{90} >$ 64 ms, and (3) for a 
variety of burst time profiles. 

BATSE's corrected integral distribution is given by 
\begin{equation}
N(>F_i) = \cases{1 & $(i = 1)$ \cr \prod_{j=2}^i (1 + \frac{1}{M_j}) & $(i > 1)$}, 
\label{21}
\end{equation}
where $F_i > F_{i+1}$, $F_i > F_{lim,i}(T_{90})$, and $M_i$ is the number of 
points in the {\it associated} set $\cal M$$_i = \{(F_j,F_{lim,j}(T_{90})) 
: F_j > F_i$ and $F_{lim,j}(T_{90}) < F_i\}$.  The corrected threshold flux, 
$F_{lim}(T_{90})$, is the minimum value of the corrected peak flux that 
satisfies the trigger criterion: $\bar F > \bar F_{lim}$, where
\begin{equation}
\bar F_{lim} = \bar C_{lim} \left(\frac {\bar F}{\bar C}\right)
\end{equation}
and $\bar C$ is the measured peak photon count rate.  By eq. \ref{20}, $F_{lim}(T_{90})$ is indeed a function of $T_{90}$ and is similarly given by
\begin{equation}
F_{lim}(T_{90}) = \bar F_{lim}(1 + \frac{\Delta t}{T_{90}}). 
\end{equation}

We apply the peak flux and integral distribution corrections of Petrosian \&
Lee (1996a) 
with one restriction:  Kouvelioutou et al. (1993), Petrosian, Lee, \& Azzam 
(1994), and Petrosian \& Lee (1996a) have demonstrated that the distribution of 
BATSE burst durations is bimodal, with the division occuring at $T_{90} \sim$ 
2 s.  This suggests that short ($T_{90} <$ 2 s) and long ($T_{90} >$ 2 s) 
duration bursts may be drawn from separate populations.  This notion is further 
supported by the tendency of short duration bursts (1) to have steeper integral 
distributions than long duration bursts (\cite{pl96a}), and (2) to have lower 
energy shifting factors than long duration bursts, especially at low peak fluxes
(\cite{mea95}).  Consequently, we exclude short duration bursts from our 
sample.

Of the 1122 bursts in the 3B catalog, information sufficient to perform these 
corrections, subject to the above restriction, exists for 423 bursts.  The 
corrected integral distribution is plotted in fig. 1.  It can be seen that the 
corrected distribution differs significantly from the uncorrected distribution 
only at peak fluxes below $F \sim$ 0.4 ph cm$^{-2}$ s$^{-1}$.  
For purposes of fitting, 
we truncate and normalize the integral distribution at $F =$ 0.316 ph cm$^{-2}$ 
s$^{-1}$, which corresponds to a trigger efficiency of $\sim \frac{1}{2}$.  The 
remaining 397 bursts are divided into eighteen bins:  fifteen are of logarithmic
length 0.1, and the brightest three are of logarithmic length 0.2.

\subsection{Time Dilation and Energy Shifting Factors}

The pulse duration time dilation factors of Norris et al. (1996a), computed 
using both peak alignment and auto-correlation statistics, are subject to energy
stretching: pulse durations tend to be shorter at higher energies 
(\cite{fea95}); consequently, pulse duration measurements of redshifted bursts 
are necessarily underestimated. Furthermore, Norris et al. (1996a) 
demonstrated that 
the unavoidable inclusion of the interpulse intervals in these analyses
has a similar effect. To correct for these effects, Norris et al. (1996a)
provided a means of calibration: they stretched 
and shifted, respectively, the time profiles and the energy spectra of the 
bursts of their reference bin by factors of 2 and 3, and from these 
``redshifted" bursts, they computed ``observed" time dilation factors.  
For each statistic, we have fitted these ``observed" time dilation factors to 
the ``actual" time dilation factors of 2 and 3 with a power law which 
necessarily passes through the origin.  Calibrated time dilation factors are 
determined from these fits and are plotted in fig. 2.

These calibrated time dilation factors are consistent with both the interpulse 
duration time dilation factors of Norris et al. (1996b) and the energy shifting 
factors (long duration bursts only) of Mallozzi et al. (1995) (see \S4), 
neither of which 
require significant energy stretching corrections.  The interpulse duration 
time dilation factors were computed for various combinations of temporal 
resolutions and signal-to-noise thresholds. Norris et al. (1996b) 
provided error 
estimates for two such combinations, which they described as ``conservative" 
with respect to their statistical significance.  These time dilation factors
and the energy shifting factors of Mallozzi et al. (1995) are additionally 
plotted in fig. 2. 
All 22 of the time dilation and energy stretching factors are fit to in \S4.

\section{Model Fits}

Both the standard candle luminosity model and the power law luminosity 
distribution model have been $\chi^2$-fitted to the corrected and binned 
differential 
distribution of fig. 1 (see \S3.1) and to the time dilation and energy 
shifting factors of fig. 2 (see \S3.2). Additionally, both models have been 
$\chi^2$-fitted to the union of these data sets.  In the case of the standard 
candle model, $\Delta\chi^2$ confidence regions, as prescribed by 
Press et al. (1989), are computed on a 100$^2$-point grid.  
In the case of the power law 
model, $\Delta\chi^2$ confidence regions are computed on a 
50$^4$-point grid and are projected into three two-dimensional planes.

\subsection{Standard Candle Luminosity Model}

The standard candle model consists of three parameters:  $h^2L_0$, $D$, 
and $z_M$, where $h = H_0/100$.  
By eqs. \ref{3c} and \ref{3d}, $z_M$ is constrained by 
\begin{equation}
z_M > \left(1 + \frac{H_0}{4c}\left(\frac{L_0}{\pi F_m}\right)^{\frac{1}{2}}\right)^2 - 1,
\label{26}
\end{equation}
where $F_m =$ 0.201 ph cm$^{-2}$ s$^{-1}$ is the peak flux of BATSE's faintest 
burst.  However, above this limit, $z_M$ is independent of the data.
 
The standard candle model fits both the differential  
distribution ($\chi^2_m =$ 18.3, $\nu =$ 16) and the time 
dilation and energy shifting factors ($\chi^2_m =$ 16.2, $\nu =$ 20).
The significance of the latter fit testifies to the consistency 
of the independent time dilation and energy shifting measurements.  The 
$\Delta\chi^2$ confidence regions of these fits (fig. 3), while demonstrating 
strong correlations between $h^2L_0$ and $D$, do not place bounds on either 
parameter. However, the latter fit places strong bounds on $h^2L_0$ for 
reasonable values of $D$.

The standard candle model additionally fits the union of these data sets 
($\chi^2_m =$ 38.2, $\nu =$ 38).  The $\Delta\chi^2$ confidence region of this 
joint fit (fig. 4) places strong bounds on both $h^2L_0$ and $D$:  $h^2L_0 =$ 
2.3$^{+0.8}_{-0.7}\times10^{57}$ ph s$^{-1}$ and 
$D =$ 3.6$^{+0.3}_{-0.3}$.  
By eq. \ref{26}, this implies that 
$z_M >$ 6.0$^{+1.5}_{-1.3}$, of which the implications are discussed in \S5.

\subsection{Power Law Luminosity Distribution Model}

The power law model consists of five parameters:  $h^2\bar L$, $D$, 
$\beta$, $K$, and $z_M$, where  
\begin{equation}
\bar L = L_m\left(\frac{1-\beta}{2-\beta}\right)\left(\frac{K^{2-\beta}-1}{K^{1-\beta}-1}\right)
\end{equation}
is the mean luminosity of the luminosity distribution, $\phi(L)$.  
The fifth parameter, $z_M$, is again constrained by 
eq. \ref{26}, except with $L_0 \rightarrow L_m$.  However, 
unlike in the standard candle model, $z_M$ is not necessarily independent of 
the data above this limit.  For purposes of fitting, we assume that $z_M$ is 
indeed beyond what BATSE observes.  The limitations of this assumption are 
discussed in \S5.

The power law model fits the differential distribution ($\chi^2_m=11.2$, 
$\nu =$ 14), the time dilation and energy shifting factors ($\chi^2_m =$ 13.6, 
$\nu =$ 18), and the union of these data sets ($\chi^2_m =$ 34.1, $\nu =$ 36).  
The $\Delta\chi^2 $ confidence region of the joint fit (fig. 5) places strong 
bounds on $D$:  $D =$ 3.7$^{+0.4}_{-0.5}$ and for $h^2\bar L <$ 10$^{57}$ 
ph s$^{-1}$, 3.4 $\la D \la$ 3.8 to 1-$\sigma$. 
This region is additionally divisible into four unique subregions (see 
tab. 1).  Using the terminology of Hakkila et al. (1995, 1996), 
the luminosity distribution of each subregion is described as $L_m$ dominated
(independent of $L_M$), $L_M$ dominated (independent of $L_m$), range 
dominated (dependent upon both $L_m$ and $L_M$), or similar to a standard candle
($L_m \sim L_M$).  For each subregion, bounds are placed on $\bar L$, $\beta$, 
$K$, and $K_{90}$, where $K_{90}$ is the 90\% width of the {\it observed} 
luminosity distribution and is given by (following the convention of Ulmer \&
Wijers (1995))
\begin{equation}
K_{90} = \frac{L_{95}}{L_5},
\end{equation}
where $L_p$, the ``$p$\% luminosity" of this distribution, is defined by
\begin{equation}
N_{L < L_{p}}(>F_m) = \left(\frac{p}{100}\right)N_{L < L_M}(>F_m).
\label{p}
\end{equation}
It is important to note that others (e.g., Horack, Emslie, \& Meegan (1994)) define $K_{90}$ differently:
\begin{equation}
K_{90} = \cases{\frac{L_{90}}{L_m} & ($L_m$ dominated) \cr \frac{L_M}{L_{10}} & ($L_M$ dominated)},
\end{equation}
which results in reduced values.  The former definition is applied here.

\section{Conclusions}

Assuming no evolution ($D =$ 3), Fenimore \& Bloom 
(1995), Nemiroff et al. (1996), and Horack, Mallozzi, \& Koshut (1996) have 
demonstrated that BATSE's differential distribution is inconsistent with a time 
dilation factor of $\sim$ 2 between the peak flux extremes of Norris et al. 
(1996a, 1996b).
This has prompted suggestions that either the bursts' observed time dilation is 
largely intrinsic or that strong evolutionary effects are present in the 
differential distribution.  The former explanation, however, is 
discredited by the degree to which the time dilation and energy shifting 
measurements are consistent.  Hakkila et al. (1996), also assuming no 
evolution, have demonstrated that the differential distribution alone is 
incompatible with a standard candle luminosity.  These results agree with our 
results for $D = 3$.  We additionally determine at what values of $D$ that
these incompatibilities disappear:  $D =$ 3.6$^{+0.3}_{-0.3}$ for the standard 
candle model and $D =$ 3.7$^{+0.4}_{-0.5}$ for the power law model.  
For mean luminosities $h^2\bar L <$ 10$^{57}$ ph s$^{-1}$, evolution 
is even more tightly constrained:  3.4 $\la D \la$ 3.8 (to 1-$\sigma$).

Horack, Emslie, \& Meegan (1994), Emslie \& Horack (1994), Ulmer \& Wijers 
(1995), Hakkila et al. (1995, 1996), and Ulmer, Wijers, \& Fenimore (1995) have 
demonstrated that $K_{90} \la$ 10 for a variety of galactic halo and 
cosmological models.  When cosmological, these models assume no evolution.  
However, when $D >$ 3, $K_{90}$ need not be so tightly constrained 
(Horack, Emslie, \& Hartmann 1995, \cite{hea_96}).  
We find that for 10$^{57}$ ph s$^{-1}$ $\la h^2\bar L \la$ 10$^{57.5}$
ph s$^{-1}$, $K_{90}$ is only constrained to be less than $\sim 10^2$ 
(see fig. 5).  Furthermore, for $h^2\bar L \la$ 10$^{56}$ ph s$^{-1}$, 
$K_{90} \ga$ 10. The former result is more conservative than estimates which 
assume no evolution.  The latter is the result of new solutions which do not 
fit the data for $D =$ 3.  

In the standard candle model, 
the redshift of BATSE's faintest burst is 6.0$^{+1.5}_{-1.3}$, which is much
greater than that which is measured for galaxies.  The power law model, under 
certain conditions, provides more reasonable estimates.  In tab. 2, 1-$\sigma$
bounds are placed on the redshift of BATSE's faintest burst for three 
representative luminosities:  $L_{10}$, $L_{50}$, and $L_{90}$, where $L_p$ is 
as defined in eq. \ref{p}. (For example, $L_{50}$ is the median luminosity of 
the {\it observed} luminosity distribution, and 80\% of the {\it observed} 
bursts have luminosities between $L_{10}$ and $L_{90}$.) 
Defining the redshift $z_p$ as the maximum redshift at which 
bursts of luminosity $L_p$ can be detected, we find that 2.9 $\la z_{50} \la$ 
4.6 for $h^2\bar L \la$ 10$^{57}$ ph s$^{-1}$ and 4.2 $\la z_{50} 
\la$ 9.4 otherwise.  However, $z_{10} \la$ 4.2 for all mean luminosities and 
$\la$ 2.3 for $h^2\bar L \la$ 10$^{57}$ ph s$^{-1}$. 
If $L_p \ga L_{90}$, the redshift of this burst is again quite large.
Consequently, a mean luminosity of $h^2\bar L \la$ 10$^{57}$ ph s$^{-1}$ 
coupled with a luminosity for BATSE's faintest burst of 
$L_p < L_{50}$ is favored.

In conclusion, the results presented in this paper demonstrate 
that when both the differential distribution and the time dilation and energy 
shifting factors are fitted to, moderate evolution is required if an 
Einstein-de Sitter cosmology, a power law spectrum of photon number index -1,
no luminosity evolution, and in the case of the power law model, a 
non-observable maximum burst redshift are assumed. We have additionally 
demonstrated that under these conditions, the 90\% width of the {\it observed} 
luminosity distribution is not necessarily $\la$ 10, as appears to be the case 
if no evolution is assumed.  Finally, redshift considerations indicate that if 
the redshifts of the faintest bursts are to be compatible with 
that which is currently known about galaxies, the standard candle model is 
unacceptable and for the power law model, a mean burst luminosity  
$h^2\bar L \la$ 10$^{57}$ ph cm$^{-2}$ s$^{-1}$ is favored.

\acknowledgments
This work was supported in part by NASA grant NAG5-2857 and an AAS/NSF-REU 
grant.  We are also grateful to E. E. Fenimore and E. D. Feigelson for 
useful discussions. 

\clearpage
\begin{deluxetable}{cccccc}
\tablecolumns{6}
\tablewidth{0pc}
\tablecaption{Power Law Model $\Delta\chi^2$ Confidence Subregions}
\tablehead{
\colhead{Subregion} & \colhead{$\phi(L)$} & \colhead{$\bar L$} & \colhead{$\beta$} & \colhead{$K$} & \colhead{$K_{90}$}} 
\startdata
1 & $L_M$ dominated & $\la L_0$ & unbounded\tablenotemark{a} & $\ga$ 10$^3$ & $\ga$ 10$^{0.5}$\tablenotemark{b} \nl
2 & range dominated & $\sim L_0$ & $\la$ 1.5 & $\la$ 10$^3$ & $\la$ 10$^2$ \nl
3 & standard candle & $\sim L_0$ & unbounded & $\sim$ 1 & $\sim$ 1 \nl
4 & $L_m$ dominated & $\sim L_0$ & $\ga$ 2.5 & $\ga$ 10$^{2.5}$ & $\la$ 10 \nl
\enddata
\tablenotetext{a}{$<$ 2 for cosmological values of $\bar L$}
\tablenotetext{b}{$<$ 10$^2$ for cosmological values of $\bar L$}
\end{deluxetable}

\clearpage
\begin{deluxetable}{ccc}
\tablecolumns{6}
\tablewidth{0pc}
\tablecaption{Power Law Model Redshift of BATES's Faintest Burst\tablenotemark{a}}
\tablehead{
\colhead{$L_p$} & \colhead{$h^2\bar L \la$ 10$^{57}$ ph s$^{-1}$} & \colhead{$h^2\bar L \ga$ 10$^{57}$ ph s$^{-1}$}} 
\startdata
$L_{10}$ & 1.0 $\la z_{10} \la$ 2.3 & 1.2 $\la z_{10} \la$ 4.2 \nl
$L_{50}$ & 2.9 $\la z_{50} \la$ 4.6 & 4.2 $\la z_{50} \la$ 9.4 \nl
$L_{90}$ & 5.1 $\la z_{90} \la$ 6.1 & 5.3 $\la z_{90} \la$ 13.1 \nl
\enddata
\tablenotetext{a}{to 1-$\sigma$}
\end{deluxetable}

\clearpage
\input psfig
\begin{figure}[htbp]
\vspace*{0 in}
\centerline{\psfig{figure=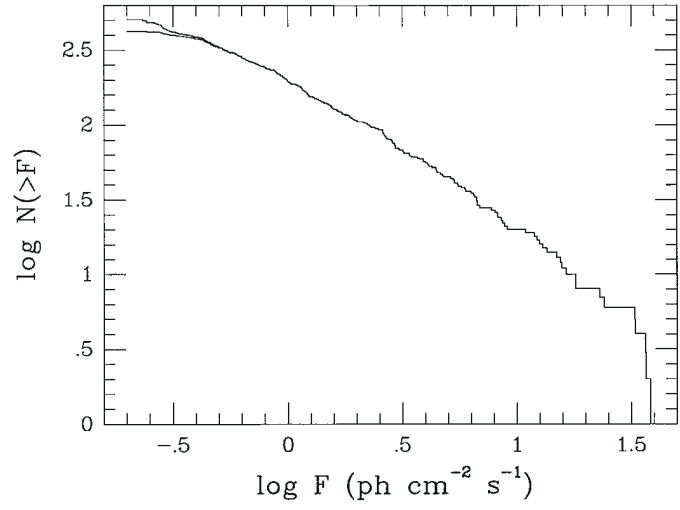,height=6.663cm,width=8.890cm}}
\caption{Uncorrected and corrected integral distributions of long duration ($T_{90} >$ 2 s) BATSE bursts.}
\end{figure}

\clearpage
\input psfig
\begin{figure}[htbp]
\vspace*{0 in}
\centerline{\psfig{figure=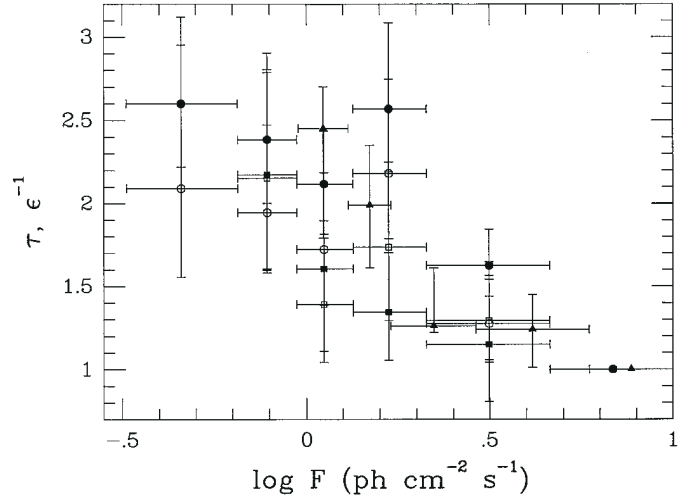,height=6.621cm,width=8.890cm}}
\caption{Calibrated (\S3.2) pulse duration time dilation factors of Norris et al. (1996a) computed using peak alignment (closed circles) and auto-correlation (open circles) statistics; interpulse duration time dilation factors of Norris et al. (1996b) computed using temporal resolutions of 512 ms (closed squares) and 128 ms (open squares) and signal-to-noise thresholds of 1400 counts s$^{-1}$ (closed squares) and 2400 counts s$^{-1}$ (open squares); and inverse peak energy shifting factors of Mallozzi et al. (1995) (closed triangles).  The time dilation factors and the energy shifting factors have been computed using two different reference bins.}
\end{figure}

\clearpage
\input psfig
\begin{figure}[htbp]
\vspace*{0 in}
\centerline{\psfig{figure=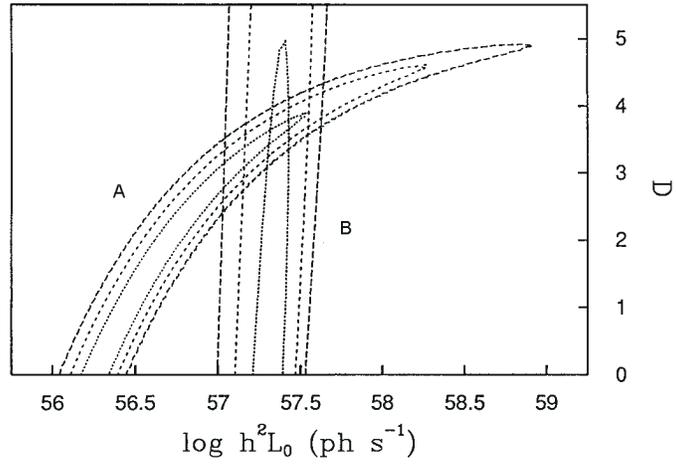,height=6.147cm,width=8.890cm}}
\caption{$\Delta\chi^2$ confidence regions of the standard candle model fit to the differential distribution (A) and to the time dilation and energy shifting factors (B).  Dotted lines are 1-$\sigma$, short dashed lines are 2-$\sigma$, and long dashed lines are 3-$\sigma$.}
\end{figure}

\clearpage
\input psfig
\begin{figure}[htbp]
\vspace*{0 in}
\centerline{\psfig{figure=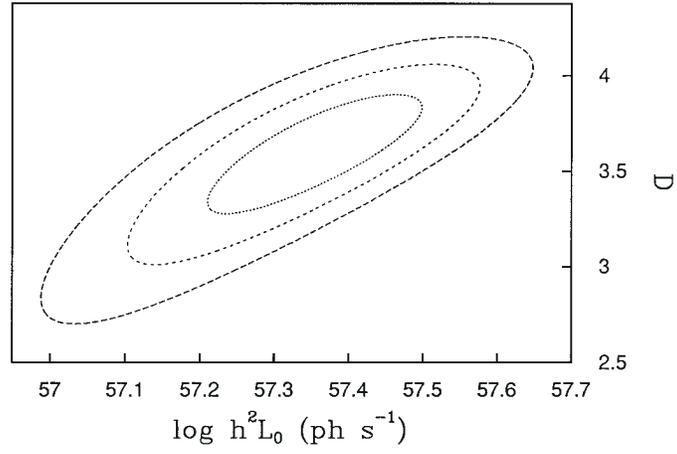,height=5.977cm,width=8.890cm}}
\caption{$\Delta\chi^2$ confidence region of the standard candle model fit to the union of the differential distribution and the time dilation and energy shifting factors.  1-, 2-, and 3-$\sigma$ are as described in fig. 3.}
\end{figure}

\clearpage
\input psfig
\begin{figure}[htbp]
\vspace*{0 in}
\centerline{\psfig{figure=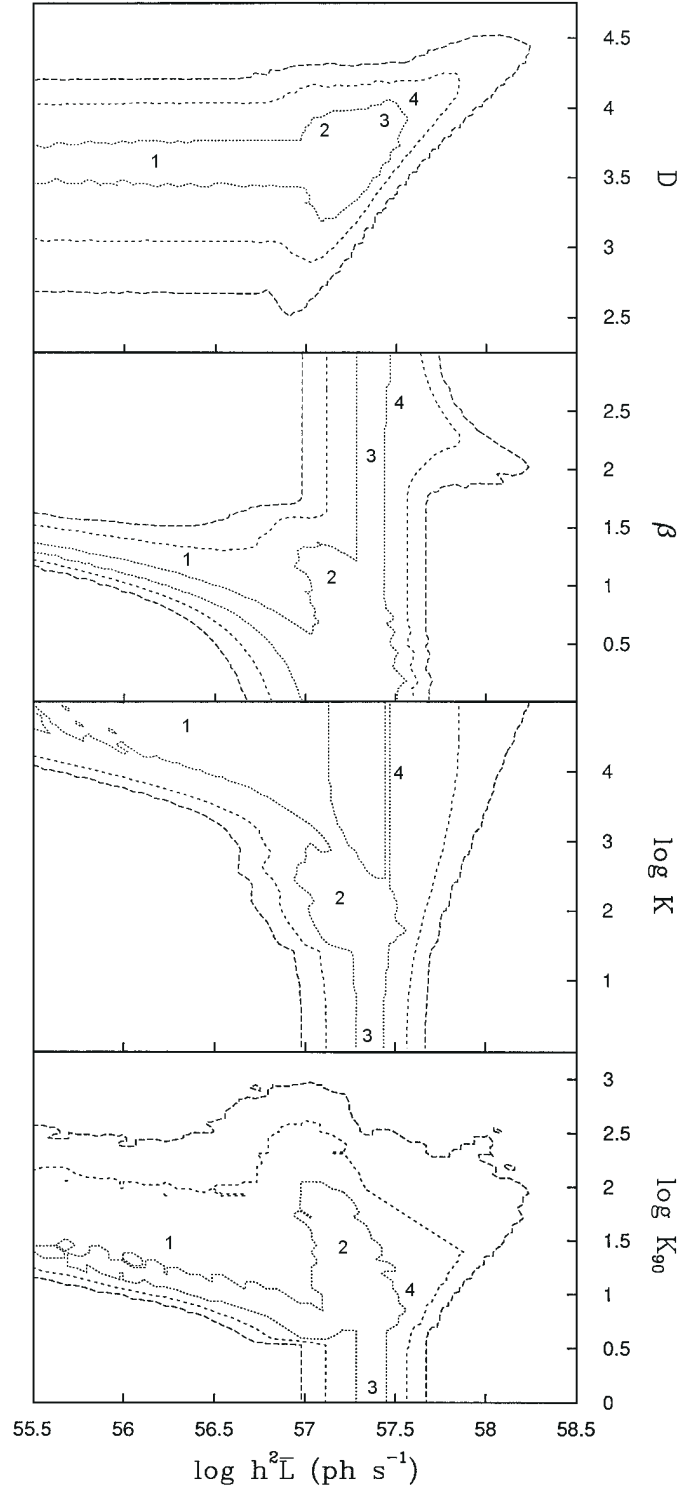,height=19.812cm,width=8.890cm}}
\caption{Projected $\Delta\chi^2$ confidence regions of the power law model fit to the union of the differential distribution and the time dilation and energy shifting factors.  1-, 2-, and 3-$\sigma$ are as described in fig. 3.  Subregions 1 - 4 are described in tab. 1.}
\end{figure}

\end{document}